\documentclass[preprint,english,showpacs,12pt,floatfix]{revtex4-1}
\usepackage{babel,amsmath,amssymb,dcolumn}
\usepackage{hyperref}
\usepackage{amsfonts}
\usepackage{amssymb}
\usepackage{graphicx}
\usepackage{latexsym}
\usepackage{dcolumn}

\begin{document}

\title{Quasinormal modes of semiclassical electrically charged black holes.}

\author{Owen Pavel Fernandez Piedra$^{1}$\footnote{Email:opavel@ucf.edu.cu}}

\affiliation{$^{1}$ Departamento de F\'{i}sica y Qu\'{i}mica,
Facultad de Mec\'{a}nica, Universidad de Cienfuegos, Carretera a
Rodas,km 4, Cuatro Caminos, Cienfuegos, Cuba}

\author{Jeferson de Oliveira$^{2}$\footnote{Email:jeferson@fma.if.usp.br}}

\affiliation{$^{2}$ Instituto de F\'{\i}sica, Universidade de S\~ao
Paulo,
  CP 66318,
05315-970, S\~ao Paulo, Brazil}

\begin{abstract}
We report the results concerning the influence of vacuum
polarization due to quantum massive vector, scalar and spinor fields
on the scalar sector of quasinormal modes in spherically symmetric
charged black holes. The vacuum polarization from quantized fields
produces a shift in the values of the quasinormal frequencies, and
correspondingly the semiclassical system becomes a better oscillator
with respect to the classical Reissner-Nordstr\"om black hole.
\end{abstract}


\maketitle

\section{Introduction}

Quantum theory and General Relativity are two cornerstones of modern
physics that for more than a century have contributed to increase
our knowledge of the Universe as never before in the human history.
With the help of the Quantum Theory we can explain micro-world
phenomena, and the General Theory of Relativity allows us a deep
understanding of the Universe at cosmological scales. Unfortunately,
this two beautiful theories resist all attempts to
bring them together. A unified theory of gravity and the quantum
world would be very important to describe, for example, the origin
of the Universe and its later development.

There are other simple phenomena that can be very interesting to
describe by the future Quantum Gravity. Among other things, from the
classical side, it is well known that the response of a black hole
to small perturbations at intermediate times is characterized, under
suitable boundary conditions, by a discrete set of complex
frequencies called quasinormal frequencies, that depend only upon
the parameters of the black hole \cite{chandra,evl,kokkotas1,nollert}. From
the quantum side, it would be interesting to see what changes appear
in the evolution of quantum black holes under perturbations. Specially
interesting is the behavior at intermediate times dominated by
quasinormal response, because apart from allowing to us to gain some
valuable information about these objects, the quasinormal spectrum permits 
investigation of the black hole stability against small
perturbations. Several numerical methods have been developed 
to study such interesting problem
\cite{price,carlemoskono,abdwangetc}.

Quasinormal modes appears to be important in other contexts, as for
example, the AdS/CFT correspondence, where the inverse of imaginary
part of quasinormal frequencies of AdS black holes can be
interpreted as the dual CFT relaxation time \cite{horowitz}
\cite{Miranda}.

In a previous paper we considered the influence of vacuum
polarizations effects due to the backreaction of a quantum massive
scalar field of large mass upon the quasinormal modes of elecrically
charged black hole solutions obtained solving the semiclassical
Einstein field equations, with the quantum renormalized stress
tensor of the quantized matter field as a source
\cite{owenjeferson1}. Such an influence appears essentially as an
appreciable shift in the quasinormal frequencies that decreases as
the bare black hole mass increases, and that not have a strong
dependance upon the quantum field parameters, leading to the
conclusion that the quantum corrected black holes are less
oscillatory with respect to its classical counterparts. Another
previous work along similar lines was done by Konoplya
\cite{konoplyabtz}, for the BTZ black hole dressed by a massless
scalar field, but in this case he considered the influence of particle
creation around the event horizon, an effect that dominates over the
vacuum polarization effect for massless fields.

To solve the backreaction problem in semiclassical gravity, we need
to know the functional dependence of the renormalized stress energy
tensor of the quantum field surrounding the classical compact object
on a wide class of metrics \cite{birrel}. Unfortunately, this is a
very difficult problem, and up to now, there exist only approximate
methods to develop a tractable expression for this quantity
\cite{birrel,Page,AHS,owen1,owenarxiv,owen2,matyjasek,owenmatyjasek1}.
Since the pioneering work of York \cite{york}, who solved the
semiclassical Einstein equations for a Schwarzschild black hole
dressed by a massless conformally coupled scalar field, using for
the quantum stress energy tensor the results given earlier by Page
\cite{Page}, there are some related works in the literature, both
for massless and massive quantum fields of different values of the
spin parameter. To see the effects of the backreaction upon the
black hole response to small perturbations, quantum massless fields
as sources of the quantum corrections are not the most suitable
candidates, because the semiclassical metric components diverge as
\(r\rightarrow\infty\) and to obtain the correct solutions to the
backreaction problem we need to impose some sort of boundary to the
system under study, a feature that causes a change the quasinormal
spectrum. A different situation happens in the case of very massive
fields, for which the vacuum polarization effects are not difficult
to compute constructing the quantum stress energy tensor by means of
the Schwinger-DeWitt expansion of the quantum effective action,
whenever the Compton's wavelenght of the field is less than the
characteristic radius of curvature
\cite{owen1,owenarxiv,matyjasek,avramidi,frolov-bv,DeWitt}.

It is important to mention that in semiclassical gravity the
unavoidable effects due to the metric fluctuations and the
associated graviton contributions to the complete quantum
stress-energy tensor are ignored, a fact that is usually justified
considering that there exists a regime in which the gravitational
field can be regarded as a classical entity, and the effects of the
remaining matter fields after quantization can be taken as quantum
corrections to the bare metric. Using the quantum stress-energy
tensor of the matter fields as a perturbation in the right hand side
of the semiclassical Einstein equations, we can obtain a
perturbative solution to the backreaction problem up to first order,
and determine what changes appear in some important quantities as
the mass, the location of the event horizon and the Hawking
temperature of the quantum corrected solution.

In this paper we study the effects that vacuum polarization of very
massive scalar, vector and spinor fields cause on quasinormal modes of quantum
corrected Reissner-Nordstr\"om black holes in four dimensions. This
is the sequel of our previous work \cite{owenjeferson1} in which we
focus on the quantum scalar field case. In the first section we
review the Schwinger-DeWitt technique to obtain the one-loop
approximation for the effective action for massive fields in the large mass limit, and present the particular results
obtained for a classical Reissner-Nordstr\"om black hole background.
In section II we solve the backreaction problem to obtain the metric
that describes the spacetime geometry of an electrically charged
semiclassical black hole. Section III is devoted to the calculation
of the massless test scalar quasinormal frequencies in this
semiclassical background, by sixth order WKB method.
Finally in Section IV we give the concluding remarks and comment
on related problems to be studied.

In the following we use for the Riemann tensor, its contractions,
and the covariant derivatives the sign conventions of Misner, Thorne
and Wheeler \cite{misner}. Our units are such that
 \(\hbar=c=G=1\).

\section{Renormalized stress energy tensor for quantum massive fields }

In the following we consider the quantization of massive scalar, vector and
spinor fields in the large mass limit. The results for the massive
scalar field can be found in our previous works \cite{owen1,owenjeferson1}, and for this reason we will be concerned only with the vector and spinor cases.
The action for a single massive vector field \(A_{\mu}\) with mass
\(m_{v}\) in some generic curved spacetime in four dimensions is
\begin{equation}\label{}
    S_{v}=-\int d^{4}x\sqrt{-g}\left(\frac{1}{4}F_{\mu\nu}F^{\mu\nu}+\frac{1}{2}m_{v}^{2}A_{\mu}A^{\mu}\right)\label{fieldaction}\quad .
\end{equation}
The equation of motion for the field have the form
\begin{equation}\label{}
    \hat{V}^{\mu}_{\nu}\left(\nabla\right)A_{\mu}=0 \label{fieldeqn1}\quad ,
\end{equation}
where the second order operator
\(\hat{V}^{\mu}_{\nu}\left(\nabla\right)\) is given by
\begin{equation}\label{}
    \hat{V}^{\mu}_{\nu}\left(\nabla\right)=\delta_{\nu}^{\mu}\Box-\nabla_{\nu}\nabla^{\mu}-R_{\nu}^{\mu}-m_{v}^{2}\delta_{\nu}^{\mu}\label{nonminimal}\quad ,
\end{equation}
where \(\Box\,=\,g^{\mu\nu}\nabla_{\mu}\nabla_{\nu}\) is the
covariant D'Alembert operator, \(\nabla_{\mu}\) is the covariant
derivative.

For single massive neutral spinor field the action is:
\begin{equation}\label{}
    S_{f}=\frac{i}{2}\int d^{4}x\sqrt{-g}\widetilde{\phi}\left[\gamma^{\mu}\nabla_{\mu}\phi+ m_{f}\phi\right]\label{class.action}\quad .
\end{equation}
In the above expression, \(\phi\) provides a spin representation of
the vierbein group and \(\widetilde{\phi}=\phi^{*}\gamma\), where *
means transpose. The Dirac matrices \(\gamma\) and\(\gamma^{\mu}\)
satisfy the usual relation
\(\left[\gamma^{\mu},\gamma^{\nu}\right]_{+}=2g_{\mu\nu}\widehat{I}\),
where \(\widehat{I}\) is the \(4\times4\) unit matrix.
\begin{widetext}
The covariant derivative of any spinor \(\zeta\) obey the
conmutation relations \cite{DeWitt,gilkey}
\begin{equation}\label{}
    \nabla_{\mu}\nabla_{\nu}\zeta-\nabla_{\nu}\nabla_{\mu}\zeta=\frac{1}{2}\mathfrak{F}_{[\alpha,\beta]}R^{\alpha\beta}_{\ \ \ \mu\nu}\quad ,
\end{equation}
\begin{equation}\label{}
    \nabla_{\nu}\nabla_{\sigma}\nabla_{\mu}\zeta-\nabla_{\sigma}\nabla_{\nu}\nabla_{\mu}\zeta=\frac{1}{2}\mathfrak{F}_{[\alpha,\beta]}R^{\alpha\beta}_{\ \ \
    \mu\sigma}\nabla_{\nu}\zeta+\nabla_{\rho}\zeta  R_{\mu \ \ \nu\sigma}^{\ \ \rho}\quad ,
\end{equation}
\begin{eqnarray}\label{}\nonumber
   \nabla_{\sigma}\nabla_{\tau}\nabla_{\nu}\nabla_{\mu}\zeta-\nabla_{\tau}\nabla_{\sigma}\nabla_{\nu}\nabla_{\mu}\zeta&=&\frac{1}{2}\mathfrak{F}_{[\alpha,\beta]}R^{\alpha\beta}_{\ \ \
    \sigma\tau}\nabla_{\nu}\nabla_{\mu}\zeta+\nabla_{\nu}\nabla_{\rho}\zeta  R_{\mu \ \ \sigma\tau}^{\ \ \rho}\\&+&\nabla_{\rho}\nabla_{\mu}\zeta  R_{\nu \ \ \sigma\tau}^{\ \ \rho}\quad ,
\end{eqnarray}
\end{widetext} and so forth, where
$\mathfrak{F}_{\left[\alpha,\beta\right]}=\frac{1}{4}\left[\gamma_{\alpha},\gamma_{\beta}\right]_{-}$
are the generators of the vierbein group, \(\left[\ \ , \ \
\right]_{-}\) is the commutator bracket, and $R^{\alpha\beta}_{\ \ \
\mu\nu}=h^{\alpha}_{\ \sigma}h^{\beta}_{\ \tau}R^{\sigma\tau}_{\ \ \
\mu\nu}$, with \(h^{\alpha}_{\ \beta}\) the vierbein which satisfies
\(h_{\alpha\mu}h^{\alpha}_{\ \nu}=g_{\mu\nu}\). The covariant
derivatives of \(\gamma\), \(\gamma^{\mu}\) and
\(\mathfrak{F}_{\left[\alpha,\beta\right]}\) vanishes. The equation
of motion for the field  \(\phi\) derived from the action
(\ref{class.action}) reads
\begin{equation}\label{}
    \left(\gamma^{\mu}\nabla_{\mu}+m_{f}\right)\phi=0 \label{diraceqn1}\quad .
\end{equation}
The operator $\hat{D}_{f}$ that gives the evolution of the spinor
function in (\ref{diraceqn1}) is:
\begin{equation}\label{}
    \hat{D}_{f}=\gamma^{\mu}\nabla_{\mu}+m_{f} \label{nonminimal2}\quad .
\end{equation}
The usual formalism of Quantum Field Theory gives an expression for
the effective action of the quantum fields \(A_{\beta}\), $\phi$ as
a perturbative expansion, 
\begin{equation}\label{}
    \Gamma\left(A_{\beta},\phi\right)=S\left(A_{\beta},\phi\right)+\sum_{k\geq1}\Gamma_{(k)}\left(A_{\beta},\phi\right)\quad ,
\end{equation}
where \(S\left(A_{\beta},\phi\right)\) is the classical action of
the free fields. The one loop contribution of the fields
\(A_{\beta}\), $\phi$ to the effective action is expressed in terms
of the operators (\ref{nonminimal}) and (\ref{nonminimal2}) as:
\begin{equation}\label{}
    \Gamma_{(1)}=\frac{i}{2}\ln\left(\mathfrak{Det}\hat{V}\right)+\frac{i}{2}\ln\left(\mathfrak{Det}\hat{D}\right)\quad ,
\end{equation}
where \(\mathfrak{Det}\hat{F}=\exp(\mathbb{T}\mathrm{r}\ln\hat{F})\)
is the functional Berezin superdeterminant of the
operator \(\hat{F}\), and \(\mathbb{T}\mathrm{r}
\hat{F}=\left(-1\right)^{i}F^{i}_{i}=\int
d^{4}x\left(-1\right)^{A}{F}^{A}_{A}(x)\) is the functional
supertrace \cite{avramidi}. If the Compton's wavelength of the field
is less than the characteristic radius of spacetime curvature
\cite{owen1,owen2,matyjasek,avramidi,frolov-bv,DeWitt,matyjasek1},
we can develope an expansion of the above effective action in powers
of the inverse square mass of the field. This is known
as the Schwinger-DeWitt approximation, and can be applied to "minimal" second
order diferential operator of the general form
\begin{equation}\label{}
     \hat{K}^{\mu}_{\nu}\left(\nabla\right)=\delta_{\nu}^{\mu}\Box-m^{2}\delta_{\nu}^{\mu}+Q^{\mu}_{\nu} \label{minimal}\quad ,
\end{equation}\
where \(Q^{\mu}_{\nu}(x)\) is some arbitrary matrix playing the role
of the potential.

Unfortunately, this is not the case of operators (\ref{nonminimal})
and (\ref{nonminimal2}). In the case of (\ref{nonminimal}), the
presence of the nondiagonal term turn it to be a nonmimal operator.
\begin{widetext}
By fortune we can put (\ref{nonminimal}) as function of some minimal
operators, if we note that it satisfies the identity
\(\hat{V}^{\mu}_{\nu}\left(\nabla\right)\left(m_{v}^{2}\delta_{\nu}^{\mu}-\nabla_{\nu}\nabla^{\mu}\right)=m_{v}^{2}\left(\delta_{\nu}^{\mu}\Box-R_{\nu}^{\mu}-m_{v}^{2}\delta_{\nu}^{\mu}\right)\)\
. Then the one loop effective action for the nonminimal operator
(\ref{nonminimal}) omitting an inessential constant can be written
as
\begin{equation}\label{}
   \frac{
   i}{2}\mathbb{T}\mathrm{r}\ln\hat{V}^{\mu}_{\nu}\left(\nabla\right)=\frac{
   i}{2}\mathbb{T}\mathrm{r}\left(\delta_{\nu}^{\mu}\Box-R_{\nu}^{\mu}-m_{v}^{2}\delta_{\nu}^{\mu}\right)-\frac{
   i}{2}\mathbb{T}\mathrm{r}\left(m_{v}^{2}\delta_{\nu}^{\mu}-\nabla_{\nu}\nabla^{\mu}\right)\label{split}\quad .
\end{equation}
\end{widetext}

We can see in (\ref{split}) that the first term is the effective
action of a minimal second order operator
\(K^{\mu}_{\nu}\left(\nabla\right)\) with potential
\(-R_{\nu}^{\mu}\). The second term can be transformed as
\(\mathbb{T}\mathrm{r}\left[\frac{1}{m_{v}^{2}}\nabla^{\mu}\nabla_{\nu}\right]^{n}=\mathbb{T}\mathrm{r}\left[\frac{1}{m_{v}^{2}}\nabla^{\mu}\Box^{n-1}\nabla_{\nu}\right]=\mathbb{T}\mathrm{r}\left[\frac{1}{m_{v}^{2}}\Box\right]^{n}\)
and \begin{equation}\label{}
    \frac{
   i}{2}\mathbb{T}\mathrm{r}\left(m_{v}^{2}\delta_{\nu}^{\mu}-\nabla_{\nu}\nabla^{\mu}\right)=\frac{
   i}{2}\mathbb{T}\mathrm{r}\left(m_{v}^{2}-\Box\right)\quad .
\end{equation}
Then, the effective action for the massive vector field is equal to
the effective action of the minimal second order operator
\(K^{\mu}_{\nu}\left(\nabla\right)\) minus the effective action of a
minimal operator \(S^{\mu}_{\nu}\left(\nabla\right)\) corresponding
to a massive scalar field minimally coupled to gravity.

The problem with the Dirac nonminimal operator \(\hat{D}_{f}\) is
solved introducing a new spinor variable \(\psi\) connected with
\(\phi\) by the relation
\(\phi=\gamma^{\sigma}\nabla_{\sigma}\psi-m_{f}\psi\) so that
(\ref{diraceqn1}) take the form
$\gamma^{\mu}\gamma^{\nu}\nabla_{\mu}\nabla_{\nu}\psi-m_{f}^{2}\psi=0$. Making use of the identity
\(\gamma^{\mu}\gamma^{\nu}\nabla_{\mu}\nabla_{\nu}=\hat{I}\left(\Box-\frac{1}{4}R\right)\) 
equation (\ref{diraceqn1}) becomes of the form
\begin{equation}\label{}
     \hat{D}_{f}^{min}\psi\equiv\left(\Box-\frac{1}{4}R-m_{f}^{2}\right)\psi=0 \label{diraceqn2}\quad ,
\end{equation}
where the potential matrix can be easily identified as
\(Q=-\frac{1}{4}R \hat{I}\).

Now using the Schwinger-DeWitt representation for the Green
functions of the minimal operators, we can obtain for the
renormalized one loop effective action of the quantum massive vector
and spinor fields the expression \(\Gamma_{(1) ren}\,=\,\int  d^{4}x
\sqrt{-g}\,\mathfrak{L}_{ren}\) where the renormalized effective
Lagrangian reads:
\begin{widetext}
\begin{equation}
\mathfrak{L}_{ren}\,=\,{1\over 2(4\pi)^{2}\,}
\sum_{k=3}^{\infty}\frac{1}{k(k-1)(k-2)}\left[{\,\mathbb{T}\mathrm{r}
\,a^{(1)}_{k}(x,x)-\,\mathbb{T}\mathrm{r} \,a^{(0)}_{k}(x,x)\over
m_{v}^{2(k-2)}}+{\mathbb{T}\mathrm{r}
\,a^{(\frac{1}{2})}_{k}(x,x)\over
m_{f}^{2(k-2)}}\right]\label{renlagrangian}\quad ,
\end{equation}
\end{widetext}

\([a^{(1)}_{k}]= \,a^{(1)}_{k}(x,x')\), \([a^{(0)}_{k}]=
\,a^{(0)}_{k}(x,x')\) and \([a^{(\frac{1}{2})}_{k}]=
\,a^{(\frac{1}{2})}_{k}(x,x')\), whose coincidence limit appears
under the supertrace operation in (\ref{renlagrangian}) are the HMDS
coefficients for the minimal operators \(\hat{K}\), \(\hat{S}\) and
$\hat{D}_{f}^{min}$ respectively. As usual, the first three
coefficients of the DeWitt-Schwinger expansion,
$a_{0},\,a_{1},\,{\rm and}\,a_{2}, $ contribute to the divergent
part of the action and can be absorbed in the classical
gravitational action by renormalization of the bare gravitational
and cosmological constants.

Restricting ourselves here to the terms proportional to
$m_{v}^{-2},$ using integration by parts and the elementary
properties of the Riemann tensor
\cite{owen1,owen2,matyjasek,owenmatyjasek1,avramidi,matyjasek1}, we
obtain for the renormalized effective lagrangian in the case of the
massive vector field considered in this work
\begin{widetext}
\begin{eqnarray}
 \nonumber
  \mathfrak{L}_{ren}&=&{1\over 192 \pi^{2} m_{v}^{2}} \left[{9\over 28} R_{\mu \nu} \Box R^{\mu \nu}- \frac{27}{280} R
 \Box R-{5\over 72} R^{3}+{31\over 60} R R_{\mu \nu } R^{\mu \nu}-{52\over 63} R^{\mu}_{\nu} R^{\nu}_{\gamma} R^{\gamma}_{\mu}
\right. \nonumber \\ &&\left. \,+\ {61\over 140} R_{\mu \nu}
R^{\mu}_{~ \sigma \gamma \varrho} R^{\nu \sigma \gamma
\varrho}-{19\over 105} R^{\mu \nu}
 R_{\gamma \varrho} R^{\gamma ~ \varrho}_{~ \mu ~ \nu}
-{67\over 2520} {R_{\gamma \varrho}}^{\mu \nu} {R_{\mu \nu}}^{\sigma
\tau} {R_{ \sigma \tau}}^{\gamma \varrho} \right. \nonumber \\
&&\left.-{1\over 10}R R_{\mu \nu \gamma \varrho} R^{\mu \nu \gamma
\varrho}\,+\,{1\over 18} R^{\gamma ~ \varrho}_{~ \mu ~ \nu} R^{\mu ~
\nu}_{~ \sigma ~ \tau} R^{\sigma ~ \tau}_{~ \gamma ~
\varrho}\right]\label{renlagrangian1}\quad .
\end{eqnarray}
\end{widetext}

The interested reader can find the general result for the spinor
field case, for example, in reference \cite{owen2}. As we can see,
this final expression of the one loop effective for the massive
vector field only differ from that of the massive scalar and spinor
fields in the numerical coefficients in front of the purely
geometric terms. For \(\langle T_{\mu\nu}\rangle_{ren}\) we obtain a
very cumbersome expression that, as in the case of
(\ref{renlagrangian1}), is different from that obtained for scalar
and spinor fields only in the numerical coefficients that appears in
front of the purely geometrical terms. For this reason we not put
this very long expression for the stress tensor here and refers the
readers to our previous papers
\cite{owen1,owen2,matyjasek,matyjasek1}.

For the present work we deal with the Reissner-Nordstr\"om
spacetime. The obtained results for the components of the
stress-tensor are very simple and can be found in reference
\cite{matyjasek}.
\begin{widetext}
For a quantum scalar field \(\phi(x)\) with mass $m$
interacting with gravity with non minimal coupling constant \(\xi\) we find
\begin{eqnarray}
\langle T_{s}\rangle^{\mu}_{\nu} &=&C^{\mu}_{\nu} +\left(\xi
-\frac{1}{6}\right)D^{\mu}_{\nu}\quad,
\end{eqnarray}
where
\begin{eqnarray}\label{}\nonumber
    C^{t}_{t}&=&-\Upsilon_{s}(1248 Q^{6} -810 r^4 Q^2 +855 M^2 r^4+202 r^2
Q^4-1878 M^3 r^3 +1152 M r^3 Q^2\\\nonumber &+&2307 M^2 r^2 Q^2 -3084 M Q^4 r),
\end{eqnarray}
\begin{eqnarray}\label{}\nonumber
    D^{t}_{t}&=&\Xi(-792 M^3 r^3 +360 M^2 r^4 +2604 M^2
Q^2 r^2-1008 M Q^2 r^3 -2712 M Q^4\\\nonumber &+&819 Q^6 +728 Q^4 r^2),
\end{eqnarray}
\begin{eqnarray}\label{}\nonumber
    C^{r}_{r}&=&\Upsilon_{s}(444 Q^6 - 1488 M Q^2 r^3 +162
Q^2 r^4 +842 Q^4 r^2-1932 M Q^4 r+315 M^2 r^4\\\nonumber &+&2127 M^2 Q^2 r^2 -462
M^3 r^3),
\end{eqnarray}
\begin{eqnarray}\label{}\nonumber
    D^{r}_{r}&=&\Xi(-792 M^3 r^3 +360 M^2 r^4 +2604
M^2 Q^2 r^2 -1008 M Q^2 r^3-2712 M Q^4 r\\\nonumber &+& 819 Q^6 +728 Q^4 r^2),
\end{eqnarray}
\begin{eqnarray}\label{}\nonumber
    C^{\theta}_{\theta}&=&-\Upsilon_{s}(3044 Q^4 r^2 -2202
M^3 r^3 -10356 M Q^4 r+3066 Q^6 -4884 M Q^2 r^3\\\nonumber &+&9909 M^2 Q^2+945
M^2 r^4 +486 Q^2 r^4 ),
\end{eqnarray}
\begin{eqnarray}\label{}\nonumber
    D^{\theta}_{\theta}&=&\Xi(3276
M^2 Q^2 r^2-1176 M Q^2 r^3 -3408 M Q^4 r +1053 Q^6-1008 M^3 r^3 \\\nonumber&+&
432 M^2 r^4 +832 Q^4 r^2),
\end{eqnarray}
For the massive vector field we obtain:
\begin{eqnarray}\label{}\nonumber
    \left\langle T_{v}\right\rangle^{t}_{t}=-\Upsilon_{v}(-31057 Q^{6} -12150 r^4 Q^2 -1665 M^2 r^4-41854 r^2
Q^4 -93537 M^2 r^2 Q^2\\\nonumber +3666 M^3 r^3 +69024 M r^3
Q^2+10751 M Q^4 r)\quad , \ \ \ \ \ \ \ \ \ \ \ \ \ \ \ \ \ \ \ \ \ \ \ \
\ \ \ \ \ \ \ \ \ \ \ \
\end{eqnarray}
\begin{eqnarray}\label{}\nonumber
    \left\langle T_{v}\right\rangle^{r}_{r}=\Upsilon_{v}( - 10448 M Q^2 r^3 +2430
Q^2 r^4 +6442 Q^4 r^2-693 M^2 r^4+12907 M^2 Q^2 r^2\\\nonumber +5365
Q^6 -16996 M Q^4 r+1050 M^3 r^3)\quad ,\ \ \ \ \ \ \ \ \ \ \ \ \ \ \ \ \ \
\ \ \ \ \ \ \ \ \ \ \ \ \ \ \ \ \ \ \ \ \ \ \ \ \ \
\end{eqnarray}
\begin{eqnarray}\label{}\nonumber
    \left\langle T_{v}\right\rangle^{\theta}_{\theta}=-\Upsilon_{v}(20908 Q^4 r^2+4854
    M^3 r^3 -44068 M Q^4 r -31708 M Q^2 r^3+7290 Q^2 r^4 \\\nonumber
    +30881
M^2 Q^2 r^2-2079 M^2 r^4 +13979 Q^6)\quad ,\ \ \ \ \ \ \ \ \ \ \ \ \ \ \ \
\ \ \ \ \ \ \ \ \ \ \ \ \ \ \ \ \ \ \ \ \ \ \
\end{eqnarray}
and for the spinor field the resulting components are given by
\begin{eqnarray}\label{}\nonumber
    \left\langle T_{f}\right\rangle^{t}_{t}=-\Upsilon_{f}(-21496\,r M{Q}^{4}+4917\,{Q}^{6}+10544\,{r}^{2}{Q}^{4}-22464\,M{r}^{3}{
Q}^{2}+21832\,{M}^{2}{r}^{2}{Q}^{2}
\\\nonumber -1080\,{M}^{2}{r}^{4}+2384\,{M}^{3}
{r}^{3}+5400\,{r}^{4}{Q}^{2})\quad , \ \ \ \ \ \ \ \ \ \ \ \ \ \ \ \ \ \ \
\ \ \ \ \ \ \ \ \ \ \ \ \ \ \ \ \
\end{eqnarray}
\begin{eqnarray}\label{}\nonumber
    \left\langle T_{f}\right\rangle^{r}_{r}=\Upsilon_{f}( -6336\,M{r}^{3}{Q}^{2}+8440\,{M}^{2}{r}^{2}{Q}^{2}+
2253\,{Q}^{6}+3560\,{r}^{2}{Q}^{4}-8680\,r M{Q}^{4}
\\\nonumber +504\,{M}^{2}{r}^{4}-784\,{M}^{3}{r}
^{3}+1080\,{r}^{4}{Q}^{2})\quad ,\ \ \ \ \ \ \ \ \ \ \ \ \ \ \ \ \ \ \ \ \
\ \ \ \ \ \ \ \ \ \ \ \ \ \ \ \ \ \ \ \ \ \ \ \ \ \
\end{eqnarray}
\begin{eqnarray}\label{}\nonumber
   \left\langle T_{f}\right\rangle^{\theta}_{\theta}=-\Upsilon_{f}(12080\,{r}^{2}{Q}^{4}-33984\,r M{e}^{4}+9933\,{Q}^{6}+30808\,{M}^{2}{r}
^{2}{Q}^{2}-20016\,M{r}^{3}{Q}^{2}
 \\\nonumber
    +1512\,{M}^{2}{r}^{4}-3536\,{M}^{3}{
r}^{3}+3240\,{r}^{4}{Q}^{2})\quad .\ \ \ \ \ \ \ \ \ \ \ \ \ \ \ \ \ \ \ \
\ \ \ \ \ \ \ \ \ \ \ \ \ \ \ \ \ \ \
\end{eqnarray}
In the above expressions, we have \(\Upsilon_{s}=\left(30240 \pi^2 m^2
r^{12}\right)^{-1}\) ,\(\Upsilon_{v}=\left(10080 \pi^2
m_{v}^2 r^{12}\right)^{-1}\),\(\Upsilon_{f}=\left(40320 \pi^2
m_{f}^2 r^{12}\right)^{-1}\) and \(\Xi=\left(720 \pi^2
m^2r^12\right)^{-1}\). $Q$ and $M$ denotes the charge and bare mass of the
black hole.
\end{widetext}
The above quantum stress energy tensors are regular at
the event horizon, as is to be expected due to the local nature of
the Schwinger-DeWitt approximation and the regular nature of the
horizon.
Also from the general form of the geometric terms
conforming the general expresion for the constructed stress tensor,
we see that it is covariantly conserved, thus indicating that it is
a good candidate for the expected exact one in our large mass
approximation.

In the specific case of a Reissner-Nordstr\"om spacetimes,
Anderson {\it{et.al}} showed, using detailed numerical results for the scalar
field case, that for \(m_{s}M\geq2\) the deviation of the
approximate stress energy tensor from the exact one lies within a
few percent \cite{AHS}. The more general condition for the validity
of the Schwinger-DeWitt approximation in the case of a spin-$j$
field can be written as \(m_{j}M\geq1\), where \(m_{j}\) and \(M\)
are respectively the field and black hole masses. In the following
we carefully take into account the above condition in numerical
calculations.

\section{Semiclassical solution}

In our previous paper we have shown how to find the general solution to
the backreaction problem in spacetimes with spherical symmetry
\cite{owenjeferson1}. In the following we solve the general
semiclassical Einstein equations assuming that there are an
electromagnetic field as a classical source, and multiple quantized
massive free fields as a perturbative quantum source, so the
solution to the backreaction problem gives a quantum corrected
Reissner-Nordstr\"om black hole.

In this limit in which we deal with free fields on a background
spacetime, the treatment of the backreaction due to a collection of
fields with different spins is easy. This is due to the fact that
the quantum stress tensors in this limit only depends quadratically
on the fields and are flavour diagonal. For example, for a set of
$N_{s}$ real scalars, upon renormalization we have
\begin{equation}\label{tmunugeneral}
    \left\langle T_{\mu}^{\nu}\right\rangle^{s}_{ren}=\sum_{k=1}^{N_{s}}\left\langle
    \left(T_{\mu}^{\nu}\right)_{k}\right\rangle^{s}_{ren}=N_{s}\left\langle \left(T_{\mu}^{\nu}\right)\right\rangle^{s}_{ren}\quad ,
\end{equation}
where $\left( T_{\mu}^{\nu}\right)$ is the classical stress tensor
for a single scalar. The last equality above follows from the fact
that the renormalization procedure is independent of the species
label $k$. In a similar manner we can arrive to the same conclusion
for the renormalization of spinor and vector field stress tensors.

The above statements permit us to obtain a good approximation to the
multiple field backreaction using as the source term in the
semiclassical Einstein equations an appropriate weighted combination
( with weights $N_{s}$, $N_{v}$ and $N_{f}$ ) of the single-species
renormalized stress energy tensors. 

In the case of our interest the general form for the line element
that solves the backreaction problem, considering only terms that
are linear in the perturbation parameter \(\varepsilon=1/M^{2}\), is
given by
\begin{equation}\label{ssmetric}
     ds^{2}=-A(r)dt^{2}+B(r)dr^{2}+r^{2}\left(d\theta^{2}+\sin^{2}\theta
     d\phi^{2}\right)\quad ,
\end{equation}
with
\begin{equation}\label{grrcomponent}
    \frac{1}{B(r)}=1-\frac{2M}{r}+\frac{Q^{2}}{r^{2}}+\frac{8 \pi}{r}\sum_{j}N_{j} \int_{\infty}^{r} \zeta^{2} \left\langle\ T_{t}^{t}
    \right\rangle_{j} d\zeta\quad ,
\end{equation}
and
\begin{equation}\label{gttcomponent1}
    A(r)=\frac{1}{B(r)}\prod_{j}\exp \left\{\lambda_{j}(r)\right\}\quad ,
\end{equation}
where
\begin{equation}\label{gttcomponent2}
    \lambda_{j}(r)=8 \pi N_{j}\int_{\infty}^{r} \zeta B\left(\zeta\right)\left(\left\langle\ T_{r}^{r}
    \right\rangle_{j}- \left\langle\ T_{t}^{t}
    \right\rangle_{j}\right)d\zeta\quad .
\end{equation}
where the subindex $j$ denotes the different single spin species
considered (scalar, vector and spinor field).
Inserting the corresponding expressions for the temporal component of the quantum stress tensor for the different fields considered in this work, we obtain
\begin{equation}\label{grrcomponentfinal1}
    \frac{1}{B(r)}=1-\frac{2M}{r}+\frac{Q^{2}}{r^{2}}+\frac{\varepsilon}{\pi}\sum_{j}\frac{N_{j}}{m_{j}^{2}}F_{j}(r)\quad,
\end{equation}
\begin{widetext}
where, for the scalar case
\begin{equation}\label{grrcomponentfinal}
    F_{s}(r)= E(r)+\xi H(r)\quad ,
\end{equation}
with
\begin{eqnarray}\label{f}\nonumber
E(r)&=&-\frac{613 M^{3}Q^4}{840 r^9} +\frac{2327 M^{2}Q^{6}}{1134
r^{10}} -\frac{3 M^{2}Q^2}{70 r^6} +\frac{5 M^{4}}{28 r^6}
-\frac{1237 M^{5} }{3780 r^{7}} +\frac{883 M^{2}Q^4}{4410
r^{8}}\\\nonumber&-&\frac{82 M^{3}Q^{^2}}{315 r^7}+\frac{1369
M^{4}Q^{2}}{1764r^8},
\end{eqnarray}
\begin{equation}\label{h}\nonumber
H(r)=\frac{28 M^{3}Q^{2}}{15 r^7}+\frac{113M^{3}Q^{4}}{30 r^{9}}
-\frac{91 M^{2}Q^{6}}{90 r^10}-\frac{52 M^{2}Q^{4}}{45 r^{8}}-
\frac{4M^{4}}{5 r^{6}} +\frac{22 M^{5}}{15 r^{7}} -\frac{62
M^{4}Q^{2}}{15 r^{8}}.
\end{equation}
\begin{eqnarray}\label{lambda}\nonumber
\lambda_{s}(r)&=&\frac{\varepsilon }{\pi m^2}\left(\frac{184M^{3}Q^{2}
}{441 r^{7}}-\frac{29 M^{4}}{140 r^{6}}-\frac{229 M^{2}Q^{4}}{840
r^{8}}+\frac{M^{2}Q^{2}}{35 r^{6}}\right)\\\nonumber&+&\frac{\varepsilon  \xi
}{\pi m^2}\left(\frac{14M^{4}}{15 r^{6}}+\frac{13 M^{2}Q^{4}}{10
r^{8}} -\frac{32 M^{3} Q^{2}}{15 r^{7}} \right).
\end{eqnarray}
and for vector and spinor fields
\begin{eqnarray}\label{f}\nonumber
F_{v}(r)&=&{\frac {26879}{2520}}\,{\frac
{{M}^{3}{Q}^{4}}{{r}^{9}}}+{\frac {2876} {315}}\,{\frac
{{M}^{3}{Q}^{2}}{{r}^{7}}}+{\frac {611}{1260}}\,{\frac
{{M}^{5}}{{r}^{7}}}-{\frac {37}{140}}\,{\frac {{M}^{4}}{{r}^{6}}}-{
\frac {20927}{4410}}\,{\frac
{{M}^{2}{Q}^{4}}{{r}^{8}}}\\\nonumber&-&{\frac {10393}
{980}}\,{\frac {{M}^{4}{Q}^{2}}{{r}^{8}}}-{\frac {27}{14}}\,{\frac
{{M }^{2}{Q}^{2}}{{r}^{6}}}-{\frac {31057}{11340}}\,{\frac
{{M}^{2}{Q}^{6} }{{r}^{10}}} \quad ,
\end{eqnarray}
\begin{eqnarray}\label{grrcomponentfinal}\nonumber
    F_{f}(r)&=& {\frac {2687}{5040}}\,{\frac {{M}^{3}{Q}^{4}}{{r}^{9}}}-{\frac {1639}{
15120}}\,{\frac {{M}^{2}{Q}^{6}}{{r}^{10}}}-{\frac {149}{1890}}\,{
\frac {{M}^{5}}{{r}^{7}}}\\\nonumber &-& \frac{3}{14}\,{\frac {{M}^
{2}{Q}^{2}}{{r}^{6}}}+{\frac {3}{70}}\,{\frac
{{M}^{4}}{{r}^{6}}}+{\frac {26}{35}}\,{\frac {{M}^{3}{Q}^{2}}{{r
}^{7}}}-{\frac {2729}{4410}}\,{\frac {{M}^{4}{Q}^{2}}{{r
}^{8}}}-{\frac {659}{2205}}\,{\frac {{M}^{2}{Q}^{4}}{{r}^{8}}}\quad ,\\
\end{eqnarray}
\begin{eqnarray}\label{lambda}\nonumber
\lambda_{v}(r)&=&\frac{\epsilon}{{\pi }{m_{v}}^{2}}\left( {\frac
{131}{420}}\,{\frac {{M}^{4}}{{r}^{6}}}+{\frac {9}{7}}\,{\frac
{{M}^{2}{Q}^{2}}{{r}^{6}}}-{\frac {9784}{2205}}\,{\frac {{M}^
{3}{Q}^{2}}{{r}^{7}}}+{\frac {2141}{840}}\,{\frac
{{M}^{2}{Q}^{4}}{{r} ^{8}}}\right)\quad  .
\end{eqnarray}
\begin{eqnarray}\label{lambda}\nonumber
\lambda_{f}(r)&=&\frac{\epsilon}{{\pi }{m_{f}}^{2}}\left( {\frac
{37}{560}}\,{\frac {{M}^{2}{Q}^{4}}{{r}^{8}}}-{\frac
{11}{210}}\,{\frac {{M}^{4}}{{r}^{6}}}+\frac{1}{7}\,{\frac
{{M}^{2}{Q}^{2}}{{r} ^{6}}}-{\frac {52}{245}}\,{\frac
{{M}^{3}{Q}^{2}}{{r}^{7}}} \right)\quad .
\end{eqnarray}
\end{widetext}
The horizon for the quantum corrected solution will be, up to first
order in the perturbation parameter, at position \(r_{+}\) given by
\begin{equation}\label{exacthorizon}
    r_{+}=r_{H}\left(1+\sum_{j}N_{j}\Lambda_{j}\right)\quad ,
\end{equation}
where
\begin{equation}\label{shifthorizon1}
    \Lambda_{j}=-\frac{4\pi}{\left(M-Q^{2}/r_{H}\right)}\int_{\infty}^{r_{h}}\zeta^{2}\left\langle T_{t}^{t}(\zeta)\right\rangle_{j}
    d\zeta \quad.
\end{equation}
and $r_{H}$ is the position of the event horizon of the bare
classical solution.
Upon substitution of the required quantities in the above expression we find
\begin{equation}\label{shifthorizonresult}
    \Lambda_{j}=\frac{\varepsilon\Gamma_{j}}{\pi
    m_{j}^{2}\left(M-Q^{2}/r_{H}\right)}\quad,
\end{equation}
with
\begin{widetext}
\begin{equation}\label{shifthorizonresult}
    \Gamma_{s}=\Theta+\xi\Omega,
\end{equation}
with
\begin{eqnarray}\label{gamma}\nonumber
\Theta&=&\frac{613 M^{3}Q^4}{1680 r_{H}^8} -\frac{2327
M^{2}Q^{6}}{22680 r_{H}^{9}} +\frac{3 M^{2}Q^2}{140 r_{H}^5}
-\frac{5 M^{4}}{56 r_{H}^6} +\frac{1237 M^{5} }{7560 r_{H}^{6}}
-\frac{883 M^{2}Q^4}{8820 r_{H}^{7}}\\\nonumber&+&\frac{41 M^{3}Q^{^2}}{315
r_{H}^6}-\frac{1369 M^{4}Q^{2}}{3528 r_{H}^7},
\end{eqnarray}
\begin{equation}\label{h}\nonumber
\Omega=-\frac{14 M^{3}Q^{2}}{15 r_{H}^6}-\frac{113M^{3}Q^{4}}{60
r_{H}^{8}} +\frac{91 M^{2}Q^{6}}{180 r_{H}^9}+\frac{26
M^{2}Q^{4}}{45 r_{H}^{7}}+ \frac{2M^{4}}{5 r_{H}^{5}} -\frac{11
M^{5}}{15 r_{H}^{6}} +\frac{31 M^{4}Q^{2}}{15 r_{H}^{7}}.
\end{equation}
and for the vector and fermion components we find
\begin{eqnarray}\label{gamma}\nonumber
\Gamma_{v}&=&{\frac {27}{28}}\,{\frac
{{M}^{2}{Q}^{2}}{{r_{{H}}}^{5}}}+{\frac {37}{ 280}}\,{\frac
{{M}^{4}}{{r_{{H}}}^{5}}}-{\frac {26879}{5040}}\,{\frac
{{M}^{3}{Q}^{4}}{{r_{{H}}}^{8}}}-{\frac {1438}{315}}\,{\frac
{{M}^{3}{ Q}^{2}}{{r_{{H}}}^{6}}}+{ \frac {31057}{22680}}\,{\frac
{{M}^{2}{Q}^{6}}{{r_{{H}}}^{9}}}\\\nonumber&-&{\frac
{611}{2520}}\,{\frac {{M}^{5}}{{r_{{H}} }^{6}}}+{\frac
{20927}{8820}}\,{\frac {{M}^{2}{Q}^{4}}{{r_{{H}}}^{7}}} +{\frac
{10393}{1960}}\,{\frac {{M}^{4}{Q}^{2}}{{r_{{H}}}^{7}}}\quad ,\\
\end{eqnarray}
\begin{eqnarray}\label{exacthorizon}\nonumber
    \Gamma_{f}&=&-{\frac {2687}{10080}}\,{\frac {{M}^{3}{Q}^{4}}{{r_{{H}}}^{7}}}+{\frac
{ 149}{3780}}\,{\frac {{M}^{5}}{{r_{{H}}}^{5}}}-{\frac
{3}{140}}\,{\frac {{M}^{ 4}}{{r_{{H}}}^{4}}}+{\frac
{3}{28}}\,{\frac {{M}^{2}{Q}^{2}}{{r_{{H}}} ^{4}}}+{\frac
{2729}{8820}}\,{\frac {{M}^{4}{Q}^{2}}{{r_{{H}}}^{6}}}\\\nonumber&-&{\frac
{13}{35}}\,{ \frac {{M}^{3}{Q}^{2}}{{r_{{H}}}^{5}}}+{ \frac
{659}{4410}}\,{\frac {{M}^{2}{Q}^{4}}{{r_{{H}}}^{6}}}+{ \frac
{1639}{30240}}\,{\frac {{M}^{2}{Q}^{6}}{{r_{{H}}}^{8}}} \quad.\\
\end{eqnarray}
\end{widetext}

\section{Scalar perturbations and quasinormal modes}


In the following we consider the evolution of a test massless scalar
field $\Phi(x^{\mu})$ with $x^{\mu}=(t,r,\theta,\phi)$, in the
quantum corrected gravitational background studied above. The
dynamics of  for this test field is governed by the Klein-Gordon
equation
\begin{equation}\label{kg1}
\frac{1}{\sqrt{-g}}\frac{\partial}{\partial
x^{\mu}}\left(\sqrt{-g}g^{\mu\nu}\frac{\partial\Phi}{\partial
x^{\nu}}\right)=0\quad,
\end{equation}
with $g_{\mu\nu}$ is the metric tensor of semiclassical solution and
$g$ its determinant. Upon separation
of the angular and radial part in the above equation and the introduction of the radial
tortoise coordinate
\begin{equation}\label{kg2}
\frac{d^2}{dr^{2}_{*}}Z_{L}-\left[\omega^2 - V\right]Z_{L}=0\quad,
\end{equation}
where $Z_{L}(r)$ denotes the radial component of the wave function,
$\omega$ is the quasinormal frequency and $V$ is the effective
potential. The potential $V$ is a function of the metric components
$g_{\mu\nu}$ and the multipolar number $L$, and for the test
massless scalar field considered in this work, is given by the
general expression
\begin{widetext}
\begin{equation}\label{potencial}
V[r(r_{*})]=A(r)\frac{L(L+1)}{r^2}+\frac{A(r)}{2rB(r)}\left[\left(\ln{A(r)}\right)'
- \left(\ln{B(r)}\right)'\right]\quad,
\end{equation}
\end{widetext}
where the prime refers to the derivative with respect to the radial
coordinate $r$. For semiclassical black holes we have in general the
following expression for the effective potential
\begin{equation}\label{}
    V(r)=V^{c}(r)+\frac{\varepsilon}{\pi}U(r) +O \left( {\epsilon}^{2}
    \right)\quad,
\end{equation}
where \(V^{c}(r)\) is the scalar effective potential of the bare
black hole solution and \(U(r)\) is a complicated function of the contributions \(\frac{N_{j}}{\,{m_{j}}^
{2}}U_{j}(r)\) due to the vacuum polarization effect related the multiple field
backreaction, that we will not write here. In the case of a classical Reissner-Nordstr\"om black
hole \(V^{c}(r)\) is given by
\begin{equation}\label{barepotential}
    V^{c}(r)={\frac { \left( {r}^{2}-2\,Mr+{Q}^{2} \right)  \left( -2\,{Q}^{2}+
\beta\,{r}^{2}+2\,Mr \right) }{{r}^{6}}}\quad,
\end{equation}
where $\beta=L(L+1)$. For the semiclassical black hole solution
considered in this paper, where the vacuum polarization effects
comes from the quantization of massive scalar, vector and spinor fields in
the large mass limit, the particular expression for \(U_{j}(r)\)
results
\begin{widetext}
\begin{equation}\label{}
    U_{s}(r)=W_{1}(r)+\xi W_{2}(r),
\end{equation}
where
\begin{eqnarray}\nonumber
W_{1}(r)=&-&{\frac {1751}{4410}}\,{\frac
{{M}^{2}{Q}^{4}}{{r}^{10}}}-{\frac {9}{20}}\,{\frac
{{M}^{4}}{{r}^{8}}}+ {\frac {1021}{540}}\,{\frac
{{M}^{5}}{{r}^{9}}}-{\frac { 1816}{945}}\,{\frac
{{M}^{6}}{{r}^{10}}}+{\frac {6}{35}} \,{\frac
{{M}^{2}{Q}^{2}}{{r}^{8}}}\\\nonumber&+&{\frac {674641}{ 158760}}\,{\frac
{{M}^{3}{Q}^{6}}{{r}^{13}}}+{\frac {17}{ 105}}\,{\frac
{{M}^{3}{Q}^{2}}{{r}^{9}}}
         -{\frac {13271}{1764}}\,{ \frac
{{M}^{4}{Q}^{4}}{{r}^{12}}}-{\frac {625}{756}}\,{\frac {{M}^{2}{
Q}^{8}}{{r}^{14}}}\\\nonumber&-&{\frac {23353}{15876}} \,{\frac
{{M}^{2}{Q}^{6}}{{r}^{12}}}+{\frac {8559}{1960}} \,{\frac
{{M}^{3}{Q}^{4}}{{r}^{11}}}-{\frac {962}{245}}\,{\frac {{M}^
{4}{Q}^{2}}{{r}^{10}}}+{\frac {16687}{2940}}\,{\frac {{M}^{5}
{Q}^{2}}{{r}^{11}}}
         \\\nonumber &+& L\left(L+1\right)\bigg(-{\frac {1529}{22680}}\,{\frac
{{M}^ {2}{Q}^{6}}{{r}^{12}}}-\frac{1}{35}\,{\frac {{M
}^{4}}{{r}^{8}}}-{\frac {1}{70}}\,{\frac {{M}^
{2}{Q}^{2}}{{r}^{8}}}\\\nonumber &+&{\frac {47}{540}}\,{\frac {{M}
^{5}}{{r}^{9}}}-{\frac {773}{17640}}\,{\frac {{M}^{2}
{Q}^{4}}{{r}^{10}}}+{\frac {44}{441}}\,{\frac {{M}^{3}
{Q}^{2}}{{r}^{9}}}+{\frac
{821}{3528}}\,{\frac {{M}^ {3}{Q}^{4}}{{r}^{11}}}-{\frac
{1171}{4410}}\,{\frac {{M}^{4}{Q}^{2}}{{r}^{10}}}\bigg),
\end{eqnarray}
and
\begin{eqnarray}\nonumber
W_{2}(r)=&&L\left(L+1\right)\bigg(-{\frac {4}{15}}\,{\frac
{{M}^{3}{Q}^{2}}{{r}^{9}}}+ {\frac {16}{15}}\,{\frac
{{M}^{4}{Q}^{2}}{{r}^{10}}} -{\frac {29}{30}} \,{\frac
{{M}^{3}{Q}^{4}}{{r}^{11}}}-{\frac {2}{5}}{\frac
{{M}^{5}}{{r}^{9}}}+{\frac {13}{90}}\,{\frac
{{M}^{2}{Q}^{4}}{{r}^{10}}}\\\nonumber&+&{\frac {13}{45}}\,{\frac
{{M}^{2}{Q}^{6}}{{r}^{12}}}+\frac{2}{15}\,{ \frac
{{M}^{4}}{{r}^{8}}}\bigg)
+ {\frac {26}{3}}\,{\frac
{{M}^{2}{Q}^{6}}{{r}^ {12}}}-{\frac {162}{5}}\,{\frac
{{M}^{5}{Q}^{2}}{{r}^{11} }}-{\frac {2101}{90}}\,{\frac
{{M}^{3}{Q}^{6}}{{r}^{13}}}\\\nonumber &+&{\frac {128}{15}}\,{\frac
{{M}^{6}}{{r}^{10}}}+2\,{\frac
{{M}^{4}}{{r}^{8}}}+\frac{13}{3}\,{\frac
{{M}^{2}{Q}^{8}}{{r}^{14}}}+{\frac {182}{ 45}}\,{\frac
{{M}^{2}{Q}^{4}}{{r}^{10}}}-{\frac {289}{10 }}\,{\frac
{{M}^{3}{Q}^{4}}{{r}^{11}}}-{\frac {28}{5}}\,{ \frac
{{M}^{3}{Q}^{2}}{{r}^{9}}}\\\nonumber &-&{\frac {42}{5}}\,{\frac
{{M}^{5}}{{r}^{9}}}+{\frac {416}{15}}\,{\frac
{{M}^{4}{Q}^{2}}{{r}^{10}}}+{\frac {130}{3}}\,{\frac
{{M}^{4}{Q}^{4}}{{r}^{12}}}.
\end{eqnarray}
for the scalar case and
\begin{eqnarray}\nonumber
U_{v}(r)=&-&{\frac {107577}{980}}\,{\frac {{M}^{5}{Q}^{2}}{{r}^{11}}
}+{\frac {306442}{2205}}\,{\frac {{M}^{4}{Q}^{2}}{{r}^{10}}}+{\frac
{34907}{196}}\,{\frac {{M}^{4}{Q}^{4}}{{r}^{12}}}+{\frac
{13}{20}}\,{\frac {{M}^{4}}{{r}^{8}}}+{\frac {54}{7}}\,{\frac
{{M}^{2}{Q}^{2}}{{r}^{8}}}\\\nonumber &-&{\frac {491}{180}}\,{\frac
{{M}^{5}}{{r}^{9}} }+{\frac {872}{315}}\,{\frac
{{M}^{6}}{{r}^{10}}}-{\frac {1205}{21}}\,{\frac
{{M}^{3}{Q}^{2}}{{r}^{9}}}-{\frac { 303071}{1960}}\,{\frac
{{M}^{3}{Q}^{4}}{{r}^{11}}}-{ \frac {15415961}{158760}}\,{\frac
{{M}^{3}{Q}^{6}}{{r}^{ 13}}}\\\nonumber &+&{\frac
{681461}{15876}}\,{\frac {{M}^{2}{Q}^{6}}{ {r}^{12}}}+ {\frac
{29027}{882}}\,{\frac {{M}^{2}{Q}^{4}}{{r}^{10}}}+{\frac {66421
}{3780}}\,{\frac
{{M}^{2}{Q}^{8}}{{r}^{14}}}+L\left(L+1\right)\bigg(- {\frac
{5}{36}}\,{\frac {{M}^{5}}{{r}^{9}}}
         \\\nonumber &+& {\frac {4678}{2205}}\,{\frac
{{M}^{3}{Q}^{2}}{{r}^{9}}}+{\frac {6653}{5880}}\,{\frac
{{M}^{3}{Q}^{4}}{{r}^{11}}}-{\frac {16067}{17640}}\,{ \frac
{{M}^{2}{Q}^{4}}{{r}^{10}}}-{\frac {4307}{22680 }}\,{\frac
{{M}^{2}{Q}^{6}}{{r}^{12}}}+\frac{1}{21}\,{ \frac
{{M}^{4}}{{r}^{8}}}\\\nonumber &-& {\frac {6257}{4410}}\,{\frac
{{M}^{4}{Q}^{2}}{{r}^ {10}}}-{\frac {9}{14}}\,{ \frac
{{M}^{2}{Q}^{2}}{{r}^{8}}}\bigg)\quad ,\\
\end{eqnarray}
and
\begin{eqnarray}\nonumber
U_{f}(r)=&-&{\frac {2279}{2520}}\,{\frac
{{M}^{2}{Q}^{8}}{{r}^{14}}}- {\frac {11101}{1470}}\,{\frac
{{M}^{5}{Q}^{2}}{{r}^{11}}} +{\frac {24764}{2205}}\,{\frac
{{M}^{4}{Q}^{2}}{{r}^{10}} }+{\frac {5092}{441}}\,{\frac
{{M}^{4}{Q}^{4}}{{r}^{12}}} -{\frac {191}{35}}\,{\frac
{{M}^{3}{Q}^{2}}{{r}^{9}}}
\\\nonumber &-&\frac{1}{10}\,{ \frac
{{M}^{4}}{{r}^{8}}}-{ \frac {119141}{21168}}\,{\frac
{{M}^{3}{Q}^{6}}{{r}^{13}} }+{\frac {113}{270}}\,{\frac {{M}
^{5}}{{r}^{9}}}+\frac{6}{7}\,{\frac
{{M}^{2}{Q}^{2}}{{r}^{8}}}-{\frac {80}{189}}\,{\frac
{{M}^{6}}{{r}^{10} }}-{\frac {8775}{784}}\,{\frac
{{M}^{3}{Q}^{4}}{{r}^{11} }}\\\nonumber &+&{\frac {23797
}{8820}}\,{\frac {{M}^{2}{Q}^{4}}{{r}^{10}}}+{\frac {3569
}{1323}}\,{\frac
{{M}^{2}{Q}^{6}}{{r}^{12}}}+L\left(L+1\right)\bigg({\frac
{7}{270}}\,{\frac {{M}^{5}}{{r}^{9}}}-{ \frac {1}{105}}\,{\frac
{{M}^{4}}{{r}^{8}}}+{\frac {12}{49}}\,{\frac {{M}^{3}{Q}^{2}}{{r}^
{9}}}\\\nonumber &-& \frac{1}{14}\,{ \frac
{{M}^{2}{Q}^{2}}{{r}^{8}}}-{\frac {3173}{35280 }}\,{\frac
{{M}^{2}{Q}^{4}}{{r}^{10}}}-{\frac {544}{2205}}\,{\frac
{{M}^{4}{Q}^{2}}{{ r}^{10}}}-{\frac {8}{189}}\,{\frac
{{M}^{2}{Q}^{6}}{{r}^{12} }}+{\frac {6659}{35280}}\,{\frac
{{M}^{3}{Q}^{4}}{{r}^{ 11}}}\bigg).\\
\end{eqnarray}
for the vector and spinor cases.
\end{widetext}

In Figure (\ref{potencial})
is presented the effective potential $V$ taking into account, as an
example, the backreaction of multiple fields with weights $N_{s}$,
$N_{v}$ and $N_{f}$ all equal to 10.

\begin{figure}[htb!] 
\begin{center}
           \includegraphics[height=8cm,width=13cm]{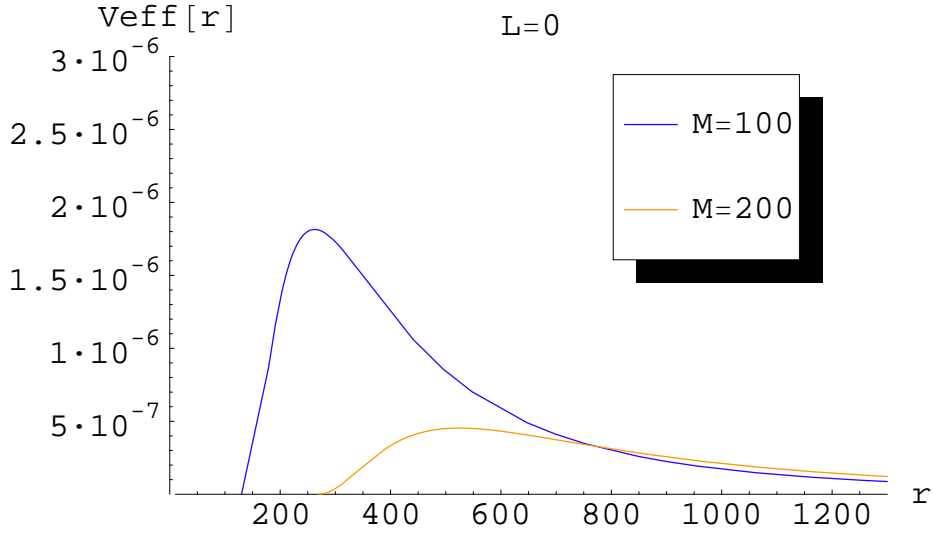}\\
           \caption{\it Effective potential of L=0 scalar modes for Semiclassical black hole with $M=100$(top curve) and $M=200$(bottom curve). $Q/M=0.95$, $N_{s}=N_{v}=N_{f}=10$ and the mass parameter of all quantum fields are chosen to be m=1/10. }
           \label{potencial}
           \end{center}
       \end{figure}
As it is observed, the figure shows a definite positive potential
barrier, i.e, a well behaved function that goes to zero at spatial
infinity and gets a maximum value near the event horizon. The
quasinormal modes are solutions of the wave equation (\ref{kg2})
with the specific boundary conditions requiring pure out-going waves
at spatial infinity and pure in-coming waves on the event horizon.
The quasinormal frequencies are in general complex numbers, whose
real part determines the real oscillation frequency and the
imaginary part determines the damping rate of the quasinormal mode.
In order to evaluate the quasinormal modes for the situation
considered in this report field, we use the well known WKB technique
at sixth order, that can give accurate values of the lowest ( that
is longer lived ) quasinormal frequencies, and  was used in several
papers for the determination of quasinormal frequencies in a variety
of systems \cite{WKB6papers}. The first order WKB technique was
applied to finding quasinormal modes for the first time by Shutz and
Will \cite{shutz-will}. Latter this approach was extended to the
third order beyond the eikonal approximation by Iyer and Will
\cite{iyer-will} and to the sixth order by Konoplya
\cite{konoplya2}.

The sixth order WKB expansion gives a relative error which is about
two orders less than the third WKB order, and allows us to determine
the quasinormal frequencies through the formula
\begin{equation}\label{WKB6}
    i\frac{\left(\omega^{2}-V_{0}\right)}{\sqrt{-2V_{0}^{''}}}-\sum_{j=2}^{6}\Pi_{j}=n+\frac{1}{2}\quad,
    \ \ \ \ \ n=0,1,2,...
\end{equation}
where \(V_{0}\) is the value of the potential at its maximum as a
function of the tortoise coordinate, and \(V_{0}^{''}\) represents
the second derivative of the potential with respect to the tortoise
coordinate at its peak. The correction terms \(\Pi_{j}\) depend on
the value of the effective potential and its derivatives ( up to the
2i-th order) in the maximum, see \cite{zhidenkothesis} and
references therein.

\begin{table}[htb!]
 { \begin{tabular}{@{}|c|c|c|c|c|c|c|c|@{}}
      \hline
      \multicolumn{4}{||c|}{Semiclassical solution}&\multicolumn{4}{|c||}{Classical solution} \\
      \hline
      \hline
      \multicolumn{8}{|c|}{$M=100$} \\
      \hline
      $L$ & $n$ & $Re(\varpi)$ & $-Im(\varpi)$ & $L$ & $n$ &
      $Re(\varpi)$  &$-Im(\varpi)$ \\
      \hline
  $ 0 $ & $0$& 10.2421  & 6.64263 &$ 0$ & $ 0 $  &5.52131  &29.4487  \\
      \hline
  $ 1 $& $0$ & 28.7831 & 6.80833  &$ 1 $&$ 0$    & 27.1215  &  7.5690 \\
      \hline
 $ 1 $& $1$ & 26.5433   & 21.0202    &$1 $ &$ 1$    & 23.2393 & 33.6626 \\
     \hline
       \hline
      \multicolumn{8}{|c|}{$M=120$}  \\
      \hline
      $L$ & $n$ & $Re(\varpi)$ & $-Im(\varpi)$ & $L$ & $n$ &
      $Re(\varpi)$  &$-Im(\varpi)$ \\
      \hline
  $ 0 $ & $0$& 8.53509 & 5.53553 &$ 0$ & $ 0 $  & 4.6011  & 24.5404 \\
      \hline
  $ 1 $& $0$ & 23.9859 &  5.6736  &$ 1 $&$ 0$    & 22.2679   &  6.30758\\
      \hline
 $ 1 $& $1$ & 22.1194   & 17.0889   &$1 $ &$ 1$    & 19.3661 & 28.0521 \\
     \hline
       \hline
      \multicolumn{8}{|c|}{$M=150$}  \\
      \hline
      $L$ & $n$ & $Re(\varpi)$ & $-Im(\varpi)$ & $L$ & $n$ &
      $Re(\varpi)$  &$-Im(\varpi)$ \\
      \hline
  $ 0 $ & $0$& 6.82809 & 4.42843 &$ 0$ & $ 0 $  & 3.68089  & 19.6324 \\
      \hline
  $ 1 $& $0$ & 19.1887  & 4.5389   &$ 1 $&$ 0$    &19.0144   & 5.04608 \\
      \hline
 $ 1 $& $1$ & 17.6956   & 14.0135     &$1 $ &$ 1$    &15.4929  & 22.4418 \\
     \hline
   \end{tabular}}
    \caption{\it Rescaled scalar quasinormal frequencies $\varpi=10^{3}\omega$ for the classical and semiclassical Reissner-Nordstr\"om black hole, with $Q/M=0.95$,$N_{s}=N_{v}=N_{f}=10$ and $m=1/10$.}\label{frequencias}
    \end{table}

The results of the numerical evaluation of the first two fundamental
quasinormal frequencies in the case considered in this work is
showed in table (\ref{frequencias}).

\begin{figure}[htb!]
\begin{center}
       \includegraphics[height=16cm,width=10cm,angle=270]{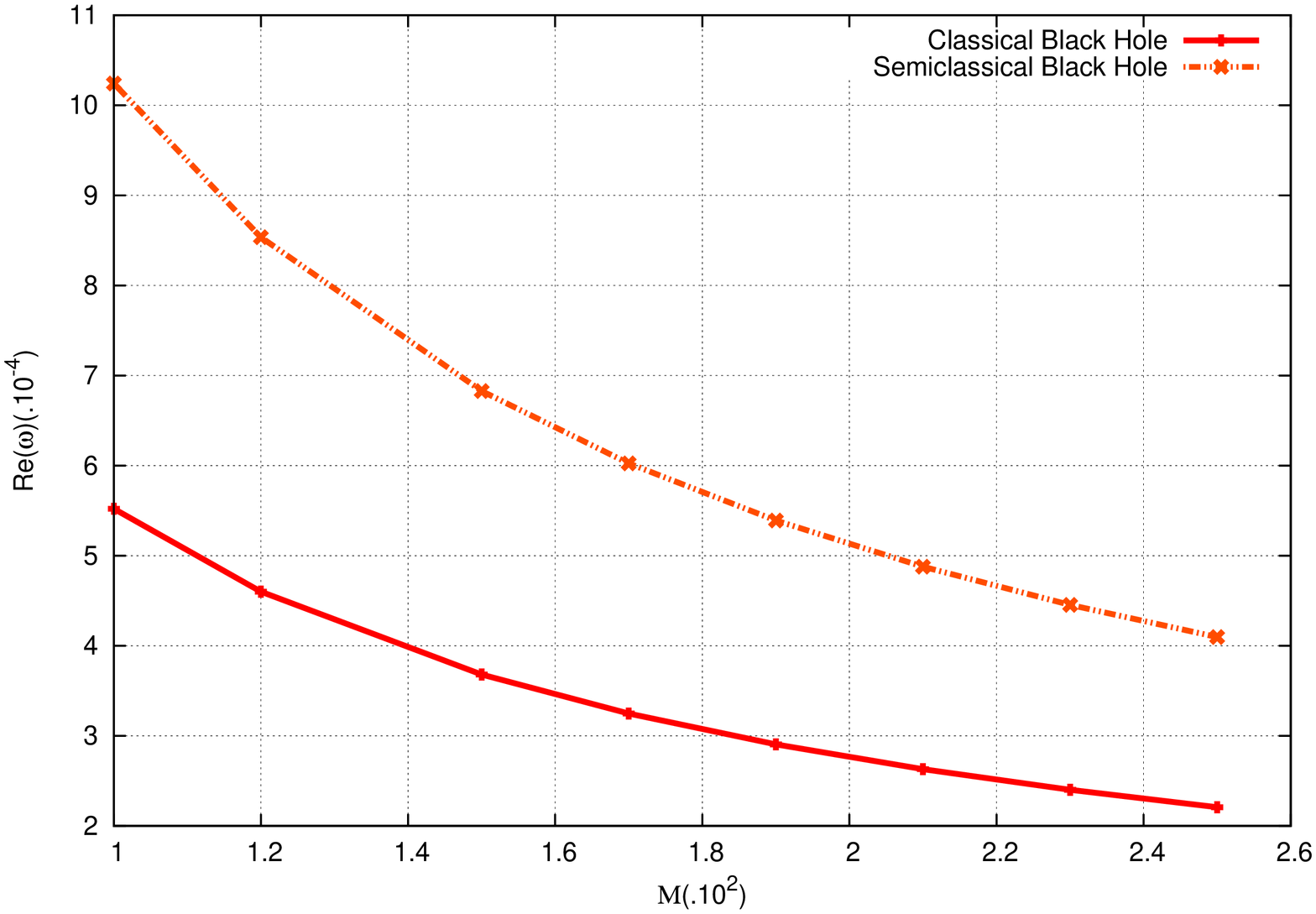}\\
           \caption{\it Dependance of $Re(\omega)$ on M for classical and semiclassical black holes. The parameters are chosen to be $Q/M=0.95$, $m=1/10$, $L=0$, $N_{s}=N_{v}=N_{f}=10$ and $n=0$.}
           \label{comparacao1}
\end{center}
\end{figure}
\begin{figure}[htb!]
\begin{center}
       \includegraphics[height=16cm,width=10cm,angle=270]{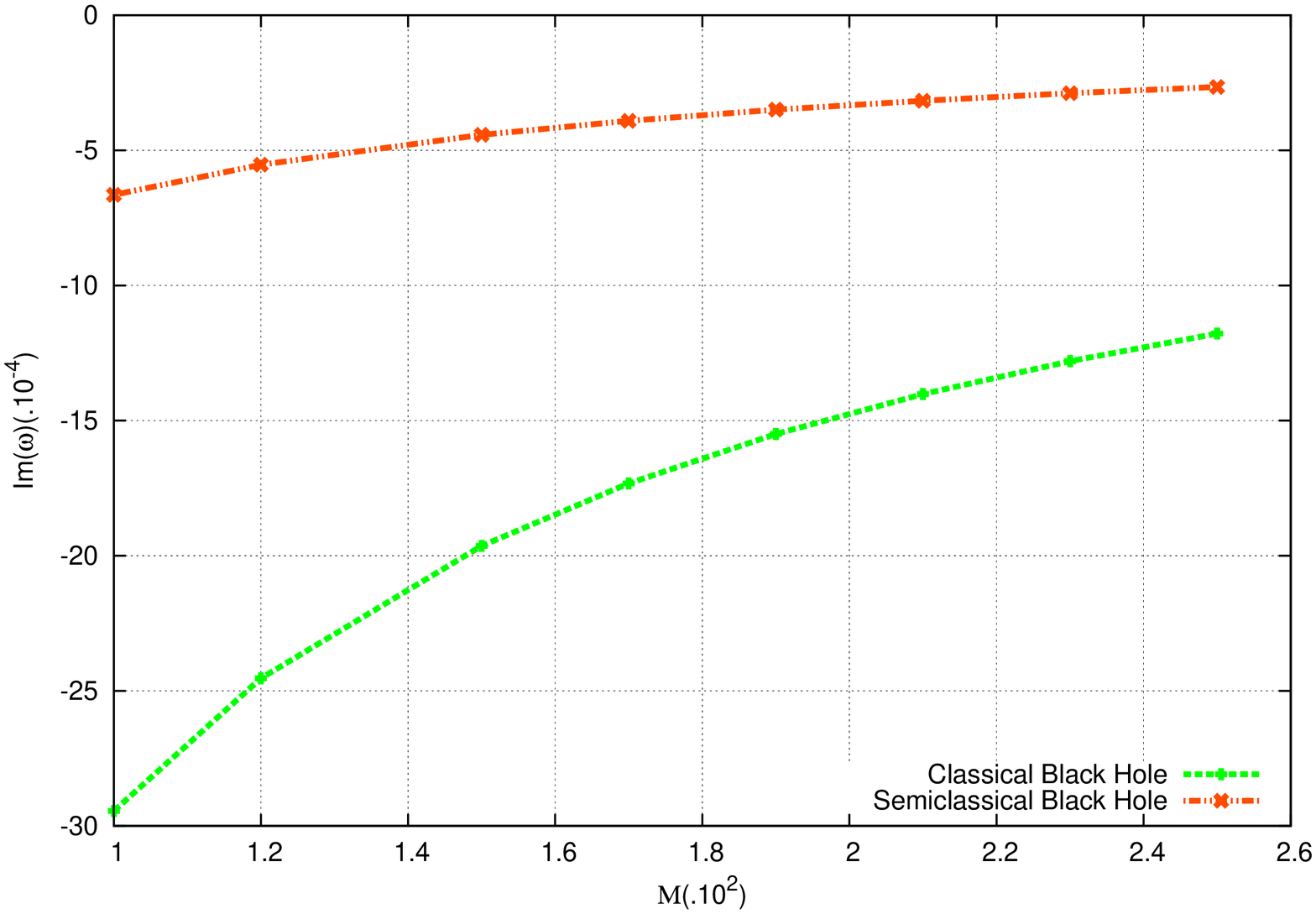}\\
           \caption{\it Dependance of $Im(\omega)$ on M for classical and semiclassical black holes. The parameters are chosen to be $Q/M=0.95$, $m=1/10$, $L=0$, $N_{s}=N_{v}=N_{f}=10$ and $n=0$.}
           \label{comparacao2}
\end{center}
\end{figure}

From figures (\ref{comparacao1}) and (\ref{comparacao2}) we see that
the backreaction of the quantized multiple fields upon the classical
charged black hole gives rise to an increasing of the real
oscillation frequencies and to a decreasing of the damping rate, for
physically interesting values of the black hole mass. Then, as a
consequence of the vacuum polarization effect due to the multiple
massive fields, we have an effective increasing of the quality
factor, proportional to the ratio
\(\frac{\left|Re(\omega)\right|}{\left|Im(\omega)\right|}\). As
expected, the differences in the quasinormal frequencies when the
black hole mass increases tend to become small. It is interesting to
note that the above results show significant differences with those
obtained previously considering only the backreaction of a quantized
massive scalar field upon the classical charged black hole solution.
In that case the quality factor is small for semiclassical black
hole with respect to the classical solution, and the shift in the
quasinormal frequencies is less pronounced. As the numerical
calculations show, this is not the case if the
separate backreaction due to massive vector and spinor fields are
considered. In each one of this cases, we obtain an increasing of
the quality factor more relevant that the decreasing showed by the
scalar field case. As a result, the combination of all types
of fields give a higher quality factor with respect to the bare
black hole. Thus, we arrive to the conclusion that the semiclassical
Reissner-Nordstr\"om black holes are better oscillators than its
classical partners. Then, in this case, there is a qualitative
correspondence with the results obtained by Konoplya in reference
\cite{konoplyabtz} for the BTZ black hole dressed by a quantum
conformal massless scalar field, where he studied the backreaction
due to Hawking radiation upon the classical spacetime.

We also studied the dependence of the quasinormal frequencies for a
fixed black hole bare mass and different values of the the quantum
field mass $m_{j}$, obtaining similar results with respect to the
massive scalar field case: a little dependance of the quasinormal
frequencies on this parameters. As the quantum field mass increases,
we found a very small increment in the real part of the frequencies
for semiclassical black holes, and a very small decrease in the
imaginary part. The same occur if we consider the variation of the
coupling constant between the massive scalar field component and the
gravitational background: the quasinormal frequencies are
insensitive to the variation of this parameter. Therefore, the shift
in the quasinormal frequencies with respect to the classical bare
black hole appears to be the same for the given range of the quantum
field masses. With respect to the multiplicity number $N_{j}$ of a
given quantum field we find that, as expected, the shift shows some
increment as this parameters increases, an effect that is more
pronounced for very large $N_{j}$. It is important to take care with
the fact that, very large values of this numbers can imply that the
total quantum stress tensor becomes a large quantity, such that the
perturbative treatment of the backreaction problem used here becomes
inadequate. By fortune, this is not the case for physically
interesting values of $N_{j}$.

\section{Concluding remarks}

We have studied the influence of the backreaction due to vacuum
polarization of multiple species of large mass quantum massive
fields, belonging to the standard model, upon the structure of
scalar quasinormal frequencies for semiclassical charged black
holes. The effect observed is a shift in the quasinormal frequencies
the semiclassical solution, such that quantum corrected black holes
becomes better oscillators that classical ones. It is important to
verify that the above effects are also true for the quasinormal
ringing phase in the evolution of spinor and electromagnetic test
fields. We are currently investigating such problems and the results
will be presented in future reports.

\section{Acknowledgements}
This work has been supported by FAPESP (\emph{Funda\c c\~ao de Amparo \`a Pesquisa do Estado de So Paulo}) and CNPQ (\emph{Conselho Nacional de Desenvolvimento Cient\' ifico e Tecnol\'ogico}), Brazil as well as
ICTP, Trieste. We are grateful to Professor Elcio Abdalla and Dr.
Alexander Zhidenko for the useful suggestions and J. Basso Marques for technical support.


\end{document}